\documentclass[runningheads,openany]{svmult}

\usepackage{makeidx}   
\usepackage{graphicx}  
\usepackage{subeqnar}
\usepackage{multicol}  
\usepackage{physprbb}
\makeindex             
\usepackage{epsfig}

\usepackage{amssymb}
\usepackage{amsfonts}
\usepackage{theorem}
\newcommand{\dir}{Figs}

\newcommand{\fig}[3]
{
     \noindent
     \unitlength=1mm
     \begin{picture}(#2,#3)
     \put(0,0){\leavevmode \epsfxsize=#2mm
       \epsffile{\dir/#1}
     }
     \end{picture}
   \vspace*{-1\baselineskip}
   \noindent
}

\newcommand{\figspace}{\vspace{1.5\baselineskip}}

\newcommand{\rr}{{\bf r}}
\newcommand{\ru}{{\bf \hat{r}}}
\newcommand{\uu}{{\bf u}}
\newcommand{\nn}{{\bf n}}

\begin{document}

\title{Spatial Order in Liquid Crystals: Computer Simulations of Systems of Ellipsoids}
\author{Friederike Schmid and Nguyen H. Phuong}
\institute{Theoretische Physik, Universit\"at Bielefeld, 
D-33501 Bielefeld, Germany}
\maketitle

\begin{abstract}
Computer simulations of simple model systems for liquid crystals
are briefly reviewed, with special emphasis on systems of ellipsoids.
First, we give an overview over some of the most commonly studied
systems (ellipsoids, Gay-Berne particles, spherocylinders).
Then we discuss the structure of the nematic phase in the bulk
and at interfaces.
\end{abstract}

\section{Introduction}

Randomly distributed hard spheres in three dimensions form two
types of structures, depending on their density: Fluid and crystalline.
For randomly distributed anisotropic particles, the situation can be different: 
Several phases may exist at intermediate densities between the fluid and the 
solid state. These phases are called ''mesophases''. For example, 
one often observes a nematic phase where the particles are oriented in one 
common preferred direction, but have no crystalline translational order. 
Other common phases are the smectic phases, in which the particles are 
arranged in layers of two dimensional fluids. Some of these structures are 
sketched in Figure 1.  Since the mesophases are neither crystalline nor 
truly liquid, they are commonly referred to as ''liquid crystal phases''.

\centerline{\fig{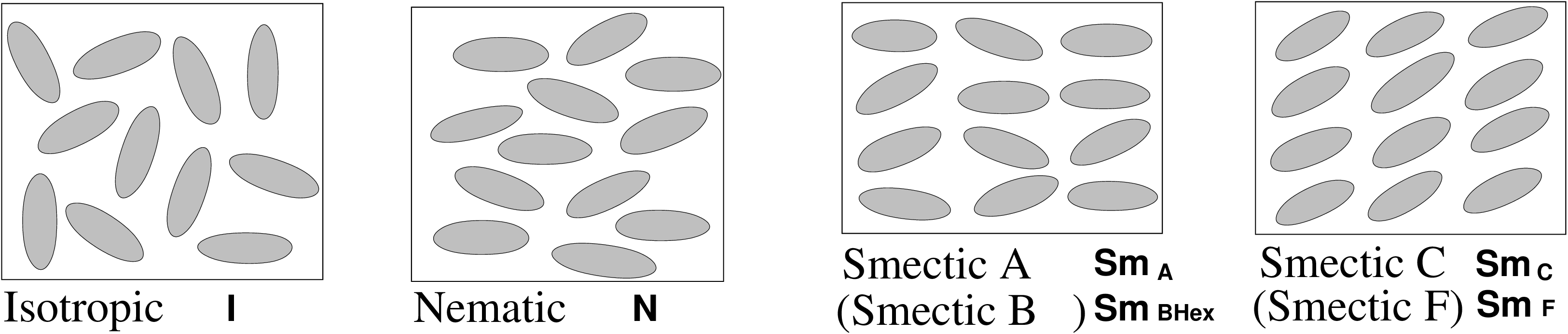}{110}{35}}
\vspace*{0.5cm}

{
\small
\noindent
{\bf Figure 1:}\\
Liquid crystalline phases. In the Smectic B${}_{\mbox{Hex}}$ and 
Smectic F phase, the structure within the layers is not entirely fluid, 
but has a type of order called hexatic. See Ref. 
\cite{degennes,chandrasekhar} 
for explanation.
}

\figspace

Experimentally, such phases are observed in various systems of anisotropic
particles: Small organic molecules, stiff polymers, self-associated
wormlike micelles, or even whole virusses (one famous example
being the tobacco mosaic virus). The phase transitions are
triggered in some cases by the temperature (thermotropic liquid
crystals), and in other cases by the concentration of the anisotropic
particles (lyotropic liquid crystals). In reality,
particles are of course not distributed ''randomly'',
but according to a Boltzmann distribution $P \propto e^{-E/k_B T}$
which accounts for the energy $E$ of their mutual interactions
($T$ is the temperature and $k_B$ the Boltzmann constant).
Nevertheless, the mechanisms which drive the phase transitions in 
lyotropic liquid crystals are essentially the same as those in systems 
of hard anisotropic particles. Therefore, the latter are often employed
to model phase transitions in liquid crystals. More generally, systems 
of anisotropic particles with simplified interactions have 
proven very useful to study generic properties of liquid 
crystals~\cite{degennes,onsager,maiersaupe,mike1,pasini}.

In the present contribution, we describe some of the computer
simulation work that has been done in this direction. We do not
intend to give a complete account of this active and rapidly 
progressing field of research. The limited space here permits only 
a rather crude introduction and the discussion of a few examples. 
The reader who is interested in more exhaustive overviews is referred 
to, e.g. , the set of review articles in Ref.~\cite{pasini}, or to
Refs.~\cite{mike1,crain,singh}.

The paper is organized as follows: First, we review some of the 
most commonly studied idealized models for particle based simulations.
We will then focus on the nematic phase and discuss some of its bulk
properties. Finally, we address specific issues of interfacial 
properties in nematic liquid crystals.

\section{Model systems}

\subsection{Ellipsoids}

Perhaps the most obvious anisotropic generalization of hard spheres 
are hard ellipsoids of revolution with one symmetry axis of length $L$
and transverse thickness $D$. The phase diagram in three dimensions
has been established from computer simulations by Frenkel, Allen and 
coworkers~\cite{frenkel1,mike2,mike3,samborski,mike4,camp}. 
It is shown for a range of elongations $\kappa = L/D$ in Figure 2.
If the particles are sufficiently anisotropic, a nematic phase (N)
intrudes between the isotropic (I) and the crystalline (X,PC) phase.
No further liquid crystalline phases are present. In particular,
smectic phases do not exist. This makes hard ellipsoids particularly 
suited to study the properties of nematic liquid crystals.

From a technical point of view, the study of ellipsoids is complicated
by the fact that the contact distance $\sigma(\uu_i,\uu_j,\ru_{ij})$
between two particles $i$ and $j$ with orientations $\uu_i$ and $\uu_j$
in the center-center direction $\ru_{ij}$ ($|\uu_{i,j}| = |\ru_{ij}|=1)$)
cannot be given in analytically closed form. Criteria to decide
whether or not two particles overlap have been derived by 
Vieillard-Baron~\cite{vieillard-baron1}, and by Perram and 
Wertheim~\cite{perram1,perram2}. Technical details on the implementation
can be found in the review article Ref.~\cite{mike3}).

\fig{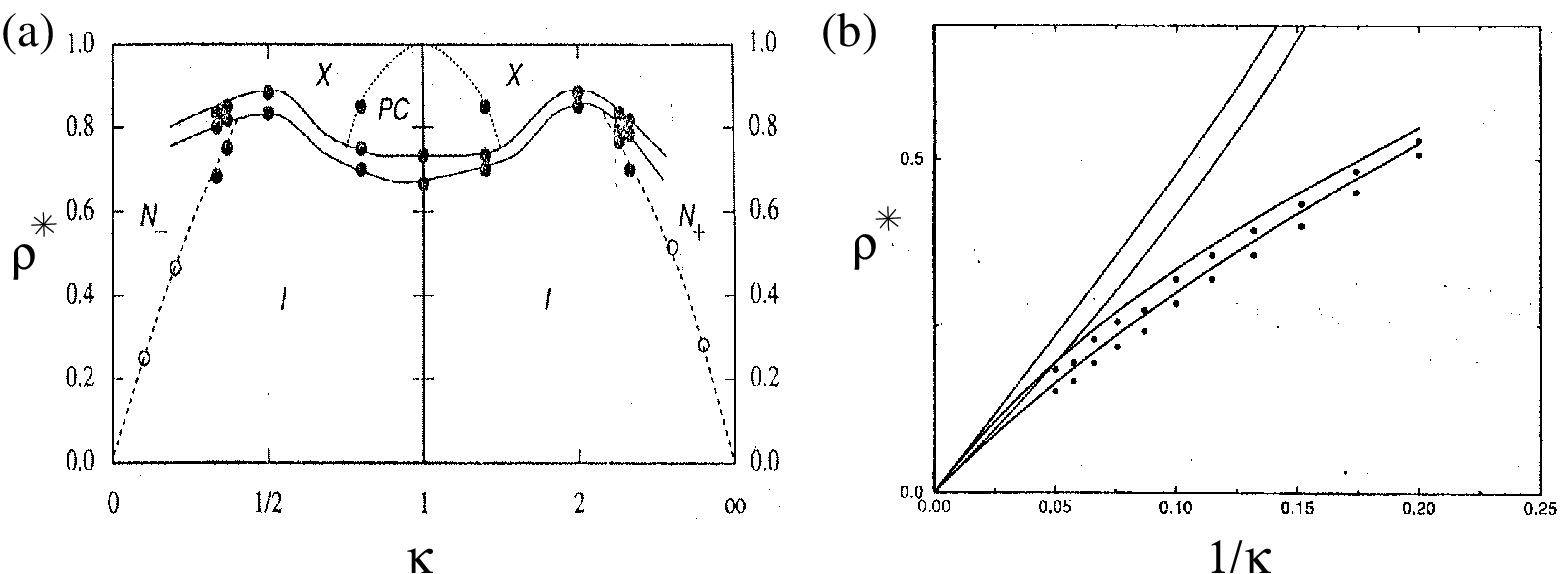}{110}{45}

{
\small
\noindent
{\bf Figure 2:}\\
Phase diagram of hard ellipsoids as a function of elongation 
$\kappa = L/D$ and reduced density $\rho/\rho_{cp}$
($\rho_{cp}$ being the density at hcp packing). 
Phases are: I (isotropic phase), N (nematic phase), 
X (orientationally ordered crystalline phase), 
PC (orientationally disordered crystalline phase).
(a) Full range of elongations: From Ref.~\cite{mike1},
    with data from Refs.~\cite{frenkel1,mike2,samborski,ladd}.
(b) Detail from Ref.~\cite{camp}. Points denote simulation data,
    lines predictions from different theories.
}

\figspace

Both the simulation and the data analysis are simplified considerably 
if an explicit formula for $\sigma(\uu_i,\uu_j,\ru_{ij})$ is available.
For this reason, Berne and Pechukas~\cite{berne} have proposed an 
approximate expression
\begin{equation}
\label{sigma}
\sigma(\uu_i,\uu_j,\ru_{ij}) =
\sigma_0 \: \left\{ 1 - \frac{\chi}{2} \left[
\frac{(\uu_i\cdot\ru_{ij}+\uu_j\cdot\ru_{ij})^2}{1+\chi \uu_i\cdot\uu_j} 
+ \frac{(\uu_i\cdot\ru_{ij}-\uu_j\cdot\ru_{ij})^2}{1-\chi \uu_i\cdot \uu_j} 
\right] \right\}^{-1/2},
\end{equation}
with the anisotropy parameter $\chi = (\kappa^2-1)/(\kappa^2+1)$
This function gives the exact contact distance 
if the particles $i$ and $j$ are co-linear (parallel to each other), 
and overestimates it by a factor of at most 
$\sqrt{2}/\sqrt{1+2 \kappa/(\kappa^2 + 1)} < \sqrt{2}$ otherwise. 
A few simulation studies have considered systems of ``hard'' particles 
with contact distances given by Eqn.~(\ref{sigma}) (``hard Gaussian overlap 
particles'')~\cite{rigby,padilla,velasco1,huang,cleaver,chrzanowska}. 
Much more commonly, however, the contact function (\ref{sigma}) is used 
in conjunction with Lennard-Jones type interaction potentials, 
the Gay-Berne potential~\cite{gay}. This class of models will be discussed 
in the next section. The present authors favor yet another potential, 
which is purely repulsive but soft.
\begin{equation}
\label{pot:se}
V_{ij} = \left\{ \begin{array}{lcr}
4 \epsilon_0 \: (R_{ij}^{-12} - R_{ij}^{-6}) + \epsilon_0,
&  & R_{ij}^{-6} < 2 \\
0, &  & \mbox{otherwise}
\end{array} \right. ,
\end{equation}
where $R_{ij}$ is a reduced distance, 
\begin{equation}
\label{reddist}
R_{ij} = (r_{ij} - \sigma(\uu_i,\uu_j,\ru_{ij}) + \sigma_0)/\sigma_0
\end{equation}
As far as we have seen so far, systems of such ``soft ellipsoids'' 
(at temperature $k_B T \sim \epsilon$) have almost the same structure
as systems of hard ellipsoids.

\centerline{\fig{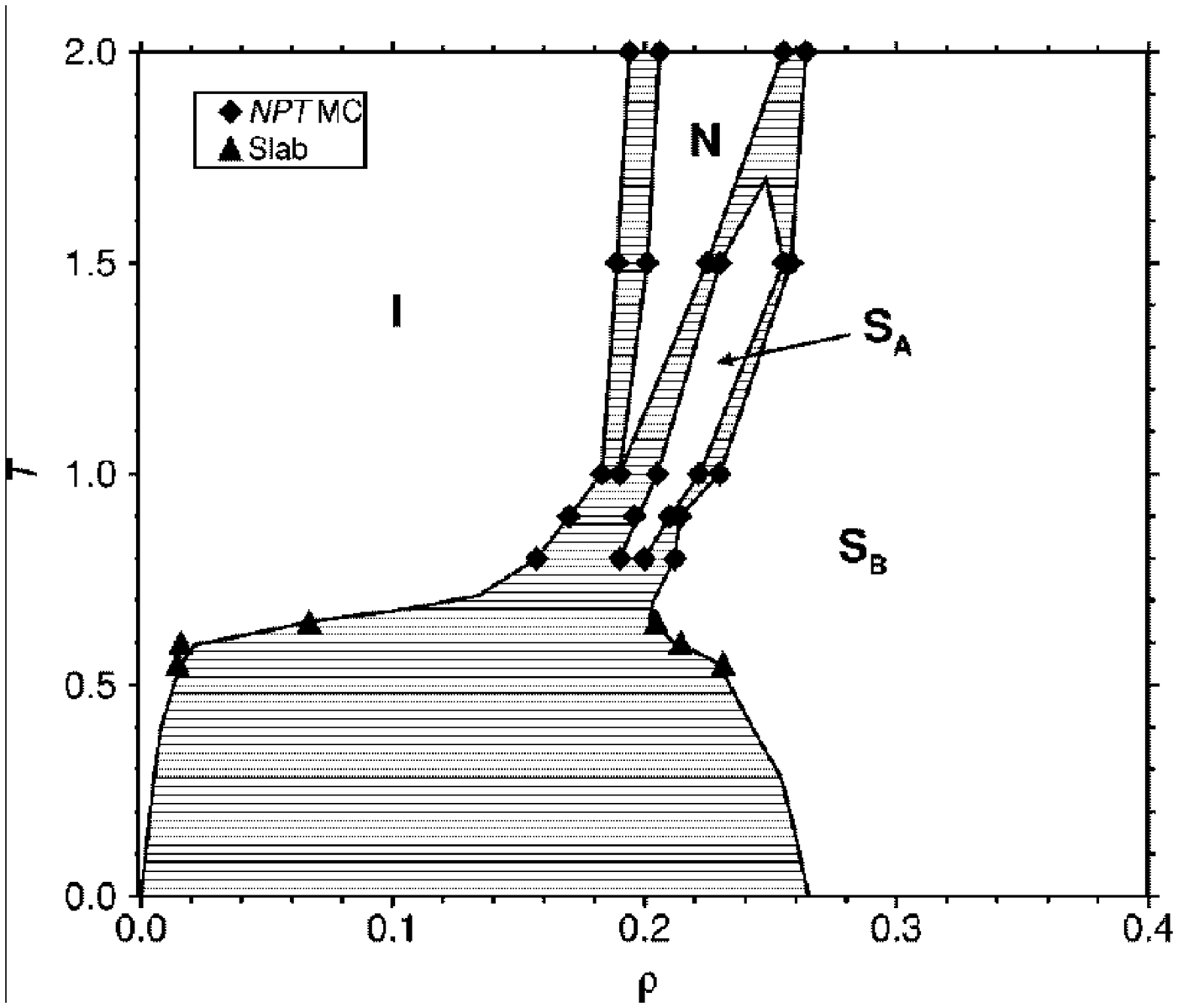}{80}{70}}

{
\small
\noindent
{\bf Figure 3:}\\
Phase diagram of Gay-Berne ellipsoids with $\mu=2$, $\nu=1$,
$\kappa=4$, and $\kappa'=5$ as a function of temperature $T$ 
in units of $\epsilon_0/k_B$ and number density $\rho$ . 
In addition to the phases observed
for hard ellipsoids (Figure 2), one also finds smectic phases 
$S_A$ and $S_B$. The simulations did not allow to decide whether
the smectic B phase is hexatic or solid, i.e., crystalline.
In the smectic A phase, however, the smectic layers are
undoubtedly fluid. From Ref.~\cite{brown}
}

\subsection{Gay-Berne particles}

Despite their appealing simplicity, hard or soft repulsive ellipsoids
are not in every respect best suited to study generic liquid crystal 
properties. For example, such models do not exhibit smectic phases.
Furthermore, one cannot study the effect of attractive interactions, 
the interplay between liquid-crystalline order and liquid-vapor phase 
separation etc. Therefore, Gay and Berne~\cite{gay} have introduced a 
class of ellipsoidal pair potentials with attractive interactions,
which has become one of the standard liquid crystal potentials.

The functional form of these potentials is
\begin{equation}
\label{pot:gb}
V_{ij} = 4 \: \epsilon_0 \: \epsilon'(\uu_i,\uu_j,\ru_{ij})^{\mu} \:
\epsilon''(\uu_i,\uu_j)^{\nu} \: (R_{ij}^{-12} - R_{ij}^{-6}),
\end{equation}
where the reduced distance $R_{ij}$ defined is as above 
(eqn. (\ref{reddist}) with $\sigma$ taken from (\ref{sigma})),
and the energy functions $\epsilon$ are given by
\begin{eqnarray}
\label{eps1}
\epsilon'(\uu_i,\uu_j,\ru_{ij})
&=& 1 - \frac{\chi'}{2} \left[
\frac{(\uu_i\cdot\ru_{ij}+\uu_j\cdot\ru_{ij})^2}{1+\chi' \uu_i\cdot\uu_j}
+ \frac{(\uu_i\cdot\ru_{ij}-\uu_j\cdot\ru_{ij})^2}{1-\chi' \uu_i\cdot
\uu_j}
\right] \\
\label{eps2}
\epsilon''(\uu_i,\uu_j) &= &
\left[ 1 - \chi^2 (\uu_i \cdot \uu_j)^2 \right]^{-1/2}.
\end{eqnarray}

The new additional anisotropy parameter $\chi'$ accounts for the fact 
that the attractive energy should favor a side-by-side alignment of particles
compared to an end-to-end alignment. It is related to the corresponding ratio 
of minimum energies (well depths) 
$\kappa' = \epsilon_{\mbox{\tiny side-side}}/\epsilon_{\mbox{\tiny end-end}}$
via $\chi' = (\kappa'^{1/\mu}-1)/(\kappa'^{1/\mu}+1)$. 
Gay and Berne originally suggested to choose the exponents $\mu=2$ and
$\nu=1$. The phase behavior for this set of exponents has been studied in 
detail~\cite{adams,demiguel1,demiguel2,emsley,demiguel3,brown},
for various choices of $\kappa$ and $\kappa'$~\cite{emsley,demiguel3,brown}.
An example is shown in Figure 3. One observes a liquid-vapor
coexistence region at low temperature. Furthermore, the attractive 
interaction stabilizes a smectic phase (Sm A), which consists of stacked
layers of two dimensional fluids. 

The Gay-Berne model is widely used in liquid crystal simulations~\cite{pasini}. 
Due to the option of varying not only $\kappa$ and $\kappa'$, but also the 
exponents $\mu$ and $\nu$, it is very versatile~\cite{luckhurst1,berardi}.
Luckhurst et al have demonstrated that it can even be adjusted to serve
as a good model for a real thermotropic liquid 
crystal~\cite{luckhurst2,bates1}.

\subsection{Other models}

Ellipsoids are not the only particles that have been used to model 
liquid crystals on an idealized level. An almost equal amount of
work has been devoted to systems of spherocylinders: Cylinders of
length $L$ and diameter $D$ capped with hemispheres at both
ends. They have the advantage that the contact distance $\sigma$
between two particles can be evaluated exactly~\cite{vieillard-baron2}.
Furthermore, they resemble quite closely certain real colloidal
liquid crystal substances, e. g., rodlike 
virusses~\cite{zasadzinski1,zasadzinski2,oldenbourg,dogic}.

The phase behavior of hard spherocylinders has been studied by Frenkel 
and others~\cite{frenkel2,frenkel3,veerman1,mcgrother,bolhuis1,polson}. 
The complete phase diagram for all elongations $L/D$ as computed by Bolhuis 
and Frenkel~\cite{bolhuis1} is shown in Figure 4. It differs strikingly 
from the corresponding phase diagram for hard ellipsoids in that sufficiently 
elongated spherocylinders form smectic phases. This came as a surprise 
when it was first discovered in simulations~\cite{frenkel2}. Meanwhile,
it has been reproduced by density functional 
theories~\cite{poniewierski1,somoza,poniewierski2,graf,velasco2}.

Bolhuis et al~\cite{bolhuis2} have studied the influence of
attractive interactions on the phase behavior of spherocylinders. 
Like in systems of Gay-Berne particles (Figure 3),
a liquid-vapor coexistence region emerges at low temperatures
(see also Ref. \cite{williamson1}).
The influence of polydispersity -- i. e., fluctuating rod length --
on the phase diagram has been investigated by Bates et al~\cite{bates2}.
Perhaps not surprisingly, the smectic phase is destabilized
in favor of the nematic phase.

We shall only briefly mention some other particle based model 
systems for liquid crystals: In a number of studies, liquid crystalline 
molecules have been modeled by stiff chain molecules
\cite{wilson,levesque1,levesque2,affouard,williamson2,alex,henning,vega}.
with aspect ratio $\kappa < 1$ and spheres cut at both 
sides have been used to study discotic liquid crystals, i. e., 
fluids of disklike particles
\cite{veerman2,bates4,deluca,emerson1,zewdie,bates5}.
Such fluids exhibit new, different types of liquid crystalline order.
For example, the smectic phases are replaced by columnar
phases: The discs assemble into hexagonal arrays of columns, but
remain fluid in one dimension within the columns.

\centerline{\fig{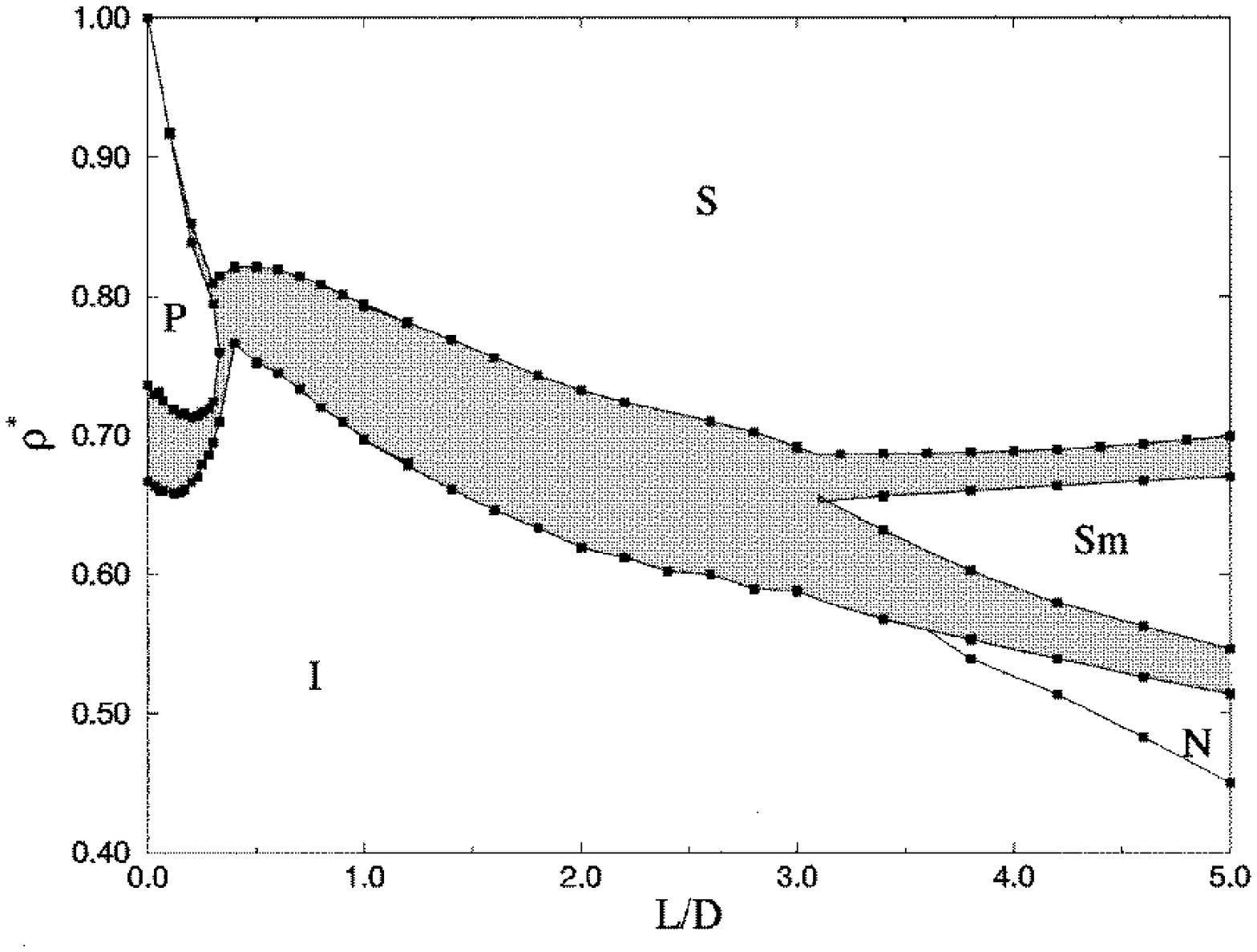}{60}{60} \fig{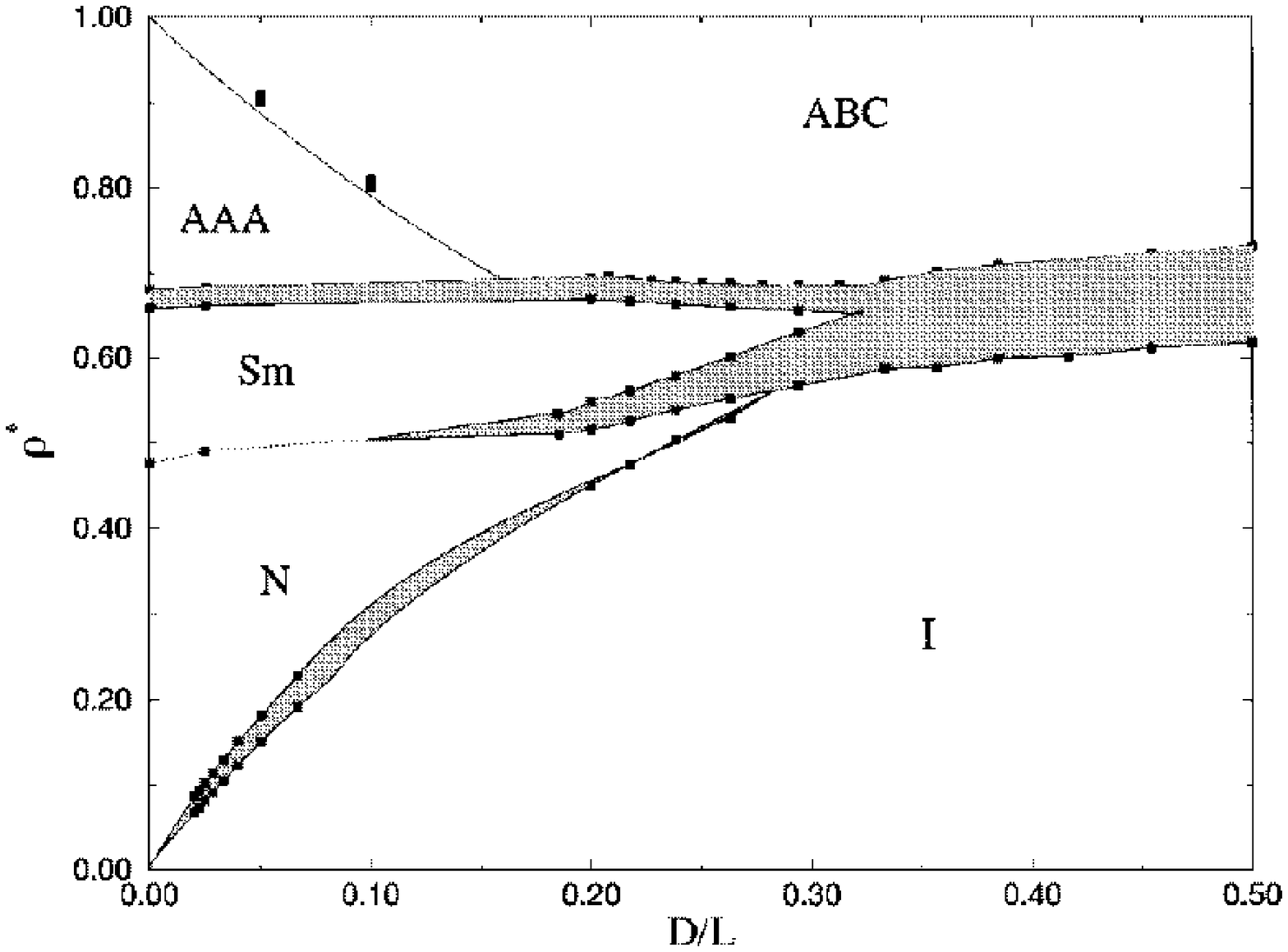}{60}{60}}

\vspace*{0.5cm}

{
\small
\noindent
{\bf Figure 4:}\\
Phase diagram of hard spherocylinders as a function of reduced 
density $\rho^* = \rho/\rho_{cp}$ and elongation $L/D$ (left)
or inverse elongation $D/L$ (right). Phases are: I (isotropic phase),
N (nematic phase), Sm (smectic A phase), AAA, S=ABC, 
(two types of solid phases with orientational order),
and P (orientationnally disordered solid phase). 
From Ref.~\cite{bolhuis1}. The transition between the nematic and 
the smectic phase is first order at all elongations~\cite{polson}.
}

\section{Properties of the nematic phase}

A number of important properties of the nematic phase can already be 
deduced from very general considerations.

First, geometric arguments suggest that the transition between the
isotropic and the nematic phase should be discontinuous or first
order, i.e., there should be a density regime in which both phases coexist. 
This is a consequence of ``Landau's rule''~\cite{degennes,landau},
which relates the character of the phase transition to the symmetry
groups of the two phases. Landau's rule is not rigorous, it does not
exclude the possibility that the width of the coexistence region
may happen to shrink to zero in certain points of the phase diagram.
Moreover, order fluctuations may affect the nature of the phase transition. 
So far, however, the actually observed isotropic-nematic transitions
have been discontinuous, albeit with rather narrow coexistence regions. 

Second, the free energy cost of spatial modulations of the average
particle orientation must vanish if the wavelength of the modulations 
tends to infinity. This is an example of Goldstone's 
theorem~\cite{goldstone1,goldstone2}: 
If the order parameter in an ordered phase breaks a continuous symmetry,
there exist soft massless fluctuation modes~\cite{forster}.
Modulations with finite wavelength are subject to elastic restoring 
forces~\cite{forster,chaikin}. For symmetry reasons, these depend on only 
three material constants, the Frank elastic constants 
$K_{11}, K_{22}, K_{33}$~\cite{oseen,zocher,frank}. Figure 5
illustrates the corresponding fundamental distortions, the ``splay'',
``twist'' and ``bend'' mode.

\centerline{\fig{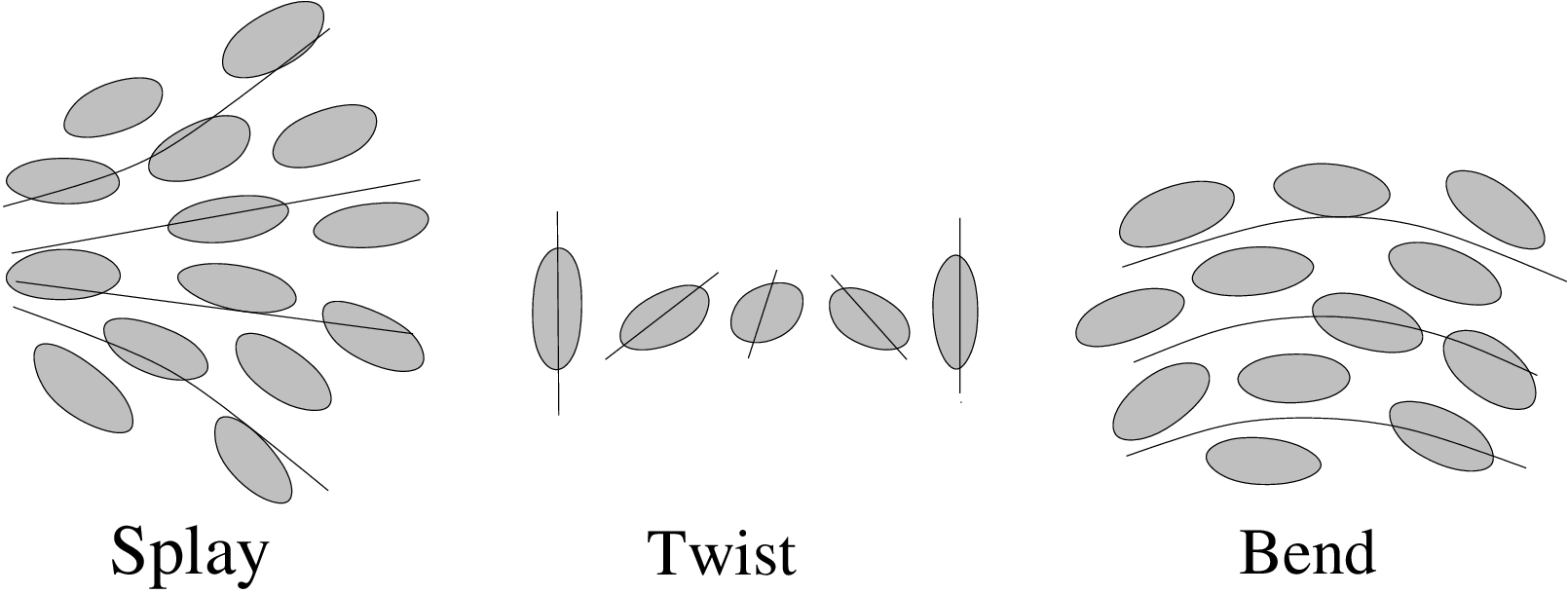}{100}{45}}

\vspace*{0.5cm}

{
\small
\noindent
{\bf Figure 5:}\\
Elastic modes in nematic liquid crystals.
}

\figspace

We turn to give a more quantitative description. The nematic order is
usually characterized by the $3\times3$ dimensional nematic order 
tensor~\cite{degennes}
\begin{equation}
\label{eq:op}
{\bf Q} = \frac{1}{N} \sum_{i=1}^{N}
(\frac{3}{2} \uu_i \otimes \uu_i - \frac{1}{2} {\bf 1} ),
\end{equation}
where the sum runs over all particles $i$, ${\bf 1}$ denotes the unit
matrix and $\otimes$ the dyadic product. The largest eigenvalue of
${\bf Q}$ is the nematic order parameter per particle, and the 
corresponding eigenvector is the director $\nn$ of the nematic liquid.
It gives the direction of preferential alignment of the molecules.
Long wavelength spatial modulations $\nn(\rr)$ of the director are
penalized by the elastic free energy~\cite{frank}
\begin{equation}
{\cal F} \{ \nn (\rr) \} = \frac{1}{2}\int d \rr \Big\{
K_{11} [ \nabla \cdot \nn ]^2 +
K_{22} [ \nn \cdot (\nabla \times \nn) ]^2 +
K_{33} [ \nn \times (\nabla \times \nn) ]^2 \Big\}.
\end{equation}
On macroscopic length scales, the structure of liquid crystals is 
controlled almost exclusively by the elastic energy and thus by the
Frank elastic constants $K_{11}$, $K_{22}$, and $K_{33}$.

The quantitative characterization of the local, microscopic structure
is much more involved. The most intensely studied quantities are usually
the pair distributions or the pair correlation functions. 
On principle, a hierarchy of infinitely many $N$-particle distribution 
functions would be required to fully characterize a fluid. In systems with 
only pairwise interactions, however, the pair distributions stand out 
because they can be used to calculate several macroscopic quantities such 
as the pressure and the compressibility. Moreover, pair correlations are
often accessible to experiments in real systems. 

As before, we characterize the orientation of a particle by a unit
vector $\uu$. The one-particle distribution $\rho^{(1)}(\uu,\rr)$
gives the probability density of finding a particle of orientation $\uu$
at the position $\rr$. Similarly, the two-particle distribution
$\rho^{(2)}(\uu_i,\rr_i,\uu_j,\rr_j)$ is the probability density of
finding simultaneously one particle of orientation $\uu_i$ at
the position $\rr_i$, and another with orientation $\uu_j$ at
the position $\rr_j$. In a homogeneous nematic phase, $\rho^{(1)}$ does not
depend on the position $\rr$, and $\rho^{(2)}$ depends only on the
relative position $\rr_{ij}=\rr_i-\rr_j$. Furthermore, particles at infinite
distances are uncorrelated, hence
\begin{equation}
\label{asym}
\rho^{(2)}(\uu_i,\uu_j,\rr) \stackrel{ r \to \infty}{\longrightarrow}
\rho^{(1)}(\uu_1) \rho^{(1)}(\uu_2).
\end{equation}
The correlations between particles at finite distance are characterized
by the total correlation function~\cite{hansen,gray}
\begin{equation}
\label{tcf}
h(\uu_1,\uu_2,\rr) = \frac{\rho^{(2)}(\uu_1,\uu_2,\rr)}
{\rho^{(1)}(\uu_{1}) \rho^{(1)}(\uu_{2})} -1,
\end{equation}
It subsumes the total effect of a particle $i$ on a 
particle $j$. Due to the elastic interactions mentioned above, it decays 
only slowly at long distances with the algebraic power law $1/r$. 

Of course, other definitions of correlation functions are possible.
As an alternative to the total correlation function, one could consider
the orientation correlation function defined by Penttinen and
Stoyan~\cite{stoyan}. It describes the pure orientation correlations 
between particles at a given distance, and should display the same
asymptotic $1/r$ power law in the nematic fluid as $h$.

The long range correlations are mediated by the whole bulk of
the surrounding nematic fluid. It seems desirable to separate these 
``indirect'' effects from a more local ``direct'' effect of two particles 
on each other. This can be done by considering the ``direct correlation 
function'' $c(\uu_i,\uu_j,\rr_{ij})$, which is defined through the 
Ornstein-Zernike
equation~\cite{hansen,gray}
\begin{equation}
\label{oz}
h({\uu_i,\uu_j,\rr_{ij}}) = c(\uu_i,\uu_j,\rr_{ij}) +
\int c(\uu_i,\uu_k,\rr_{ik}) \: \rho^{(1)}(\uu_k) \:
 h(\uu_k,\uu_j,\rr_{kj}) d\uu_{k} d\rr_{k}.
\end{equation}
As it turns out, the direct correlation function is indeed short
range even in the nematic phase. 

We illustrate this with recent computer simulation results.
We have performed computer simulations of 1000-8000 ellipsoids
with elongation $\kappa=3$ at the number density $\rho = 0.3$, 
interacting with the potential (\ref{pot:se}) at temperature $T=0.5$.
Technical details can be found in Refs.~\cite{phuong1,phuong2,phuong3}. 

The calculation of direct correlation functions from computer 
simulations is computationally demanding and involves a complicated data 
analysis. We were the first to determine the direct correlation function 
from computer simulations of a nematic fluid without any approximations. 
Direct correlation functions in isotropic fluids have been calculated 
earlier by Allen et al~\cite{mike6,mike7}, and approximate data for nematic 
fluids have been derived from simulations by Stelzer et al~\cite{stelzer1,stelzer2}.

Figure 6 shows the orientational average of the total correlation
function and the direct correlation function. The average is
carried out over the orientations $\uu_i, \uu_j$, and $\ru_{ij}$.
Figure 6 demonstrates that long range algebraic correlations between 
particles are not apparent in this averaged curve. The elastic interactions 
affect only orientations, not total densities. Therefore, both $h$ and $c$ 
are short range.

\centerline{\fig{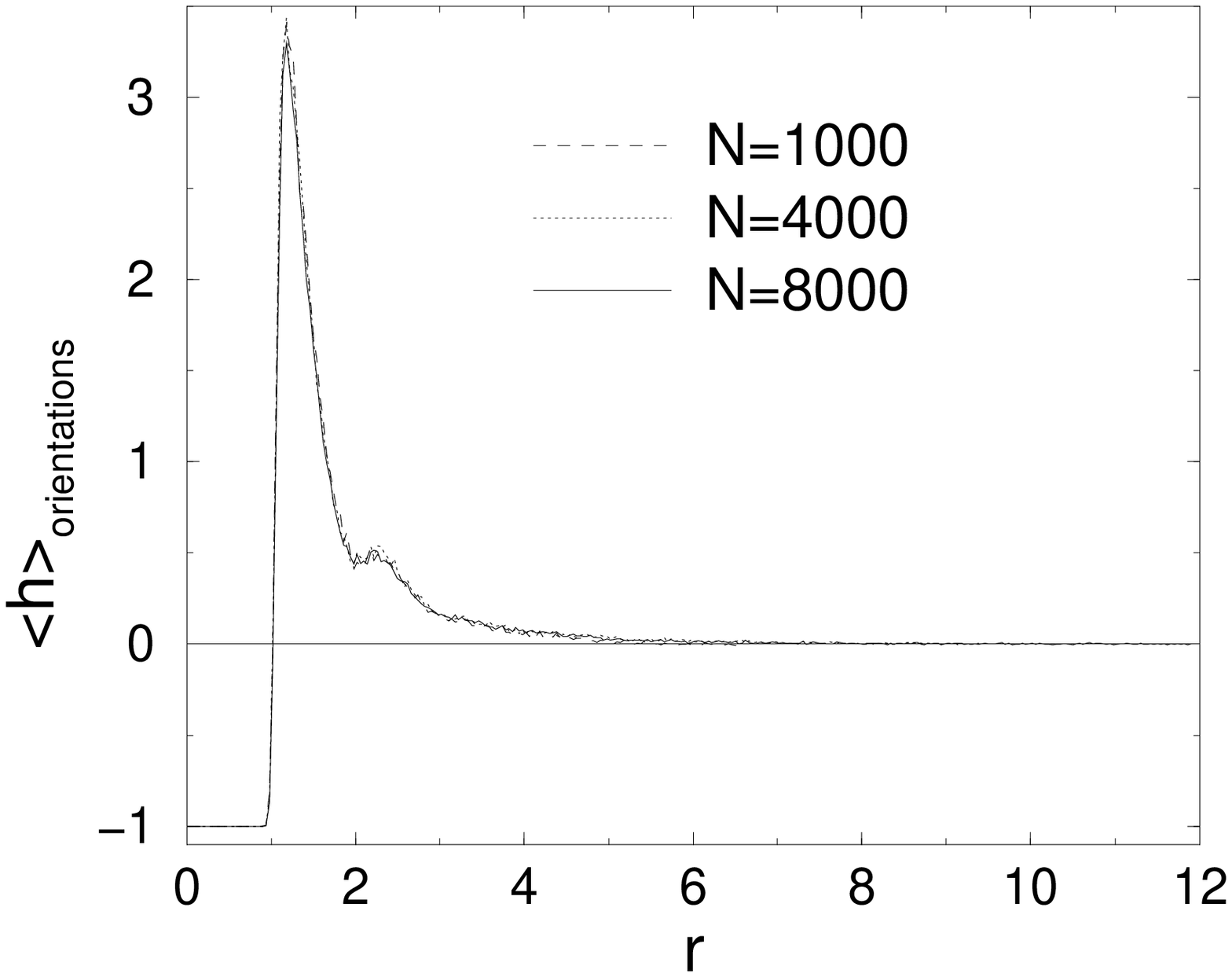}{55}{55} \fig{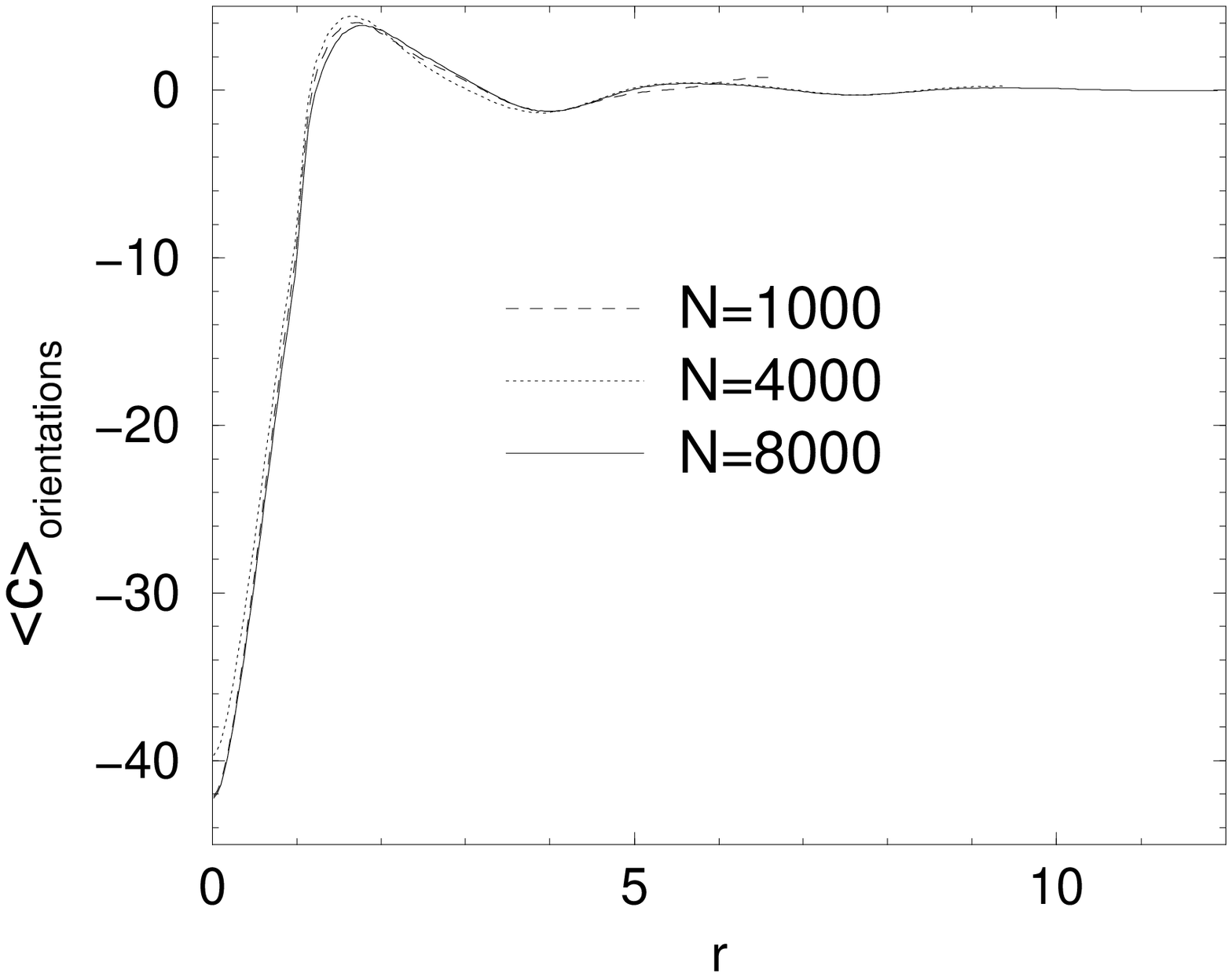}{55}{55}}

\vspace*{0.5cm}

{
\small
\noindent
{\bf Figure 6:}\\
Total correlation function $h$ (left) and direct correlation function
$c$ (right) vs. molecular distance $r$, averaged over all orientations 
of $\uu_i$,$\uu_j$, and $\rr_{ij}$, for different system sizes $N$.
}

\figspace

In order to assess the effect of the elasticity, we need to study
orientational dependent correlations. To this end, we expand all
correlation functions in spherical harmonics $Y_{lm}(\uu)$.
\begin{equation}
F(\uu_i,\uu_j,\rr) =
\sum_{{{l_1,l_2,l}\atop {m_1,m_2,m}}} \!\!
F_{l_1 m_1 l_2 m_2 l m}(r) \:
Y_{l_{1}m_{1}}(\uu_i) \: Y_{l_{2}m_{2}} (\uu_j) \: Y_{lm}(\ru),
\end{equation}
where $F$ stands for $\rho^{(2)}$, $h$, or $c$. The $z$ axis is chosen 
in the direction of the director. The orientation dependence of the 
correlations is reflected by the coefficients with nonzero indices. 
Figure 7 (left) shows a coefficient of the total correlation function 
with a particularly pronounced long range tail. As shown in 
Figure 7 (right), the tail completely disappears in the  
direct correlation function.

The direct correlation function is a central quantity in theories of
liquid matter~\cite{hansen}. It enters density functional theories
which predict the structure of inhomogeneous fluids, e. g., fluids at
interfaces. It allows to calculate a number of material constants of 
fluids. In nematic fluids, for example, it can be used to calculate
the Frank elastic constants according to a set of equations first derived 
by Poniewierski and Stecki~\cite{poniewierski3,poniewierski4}. 
We have applied these equations to our data and obtained a set of 
values for $K_{11}, K_{22}$, and $K_{33}$, which is in agreement 
with results from a different, more direct method~\cite{phuong1,mike5}. 
Thus the direct correlation function establishes a bridge between 
the microscopic structure and the parameters which characterize the 
macroscopic structure. 

\centerline{\fig{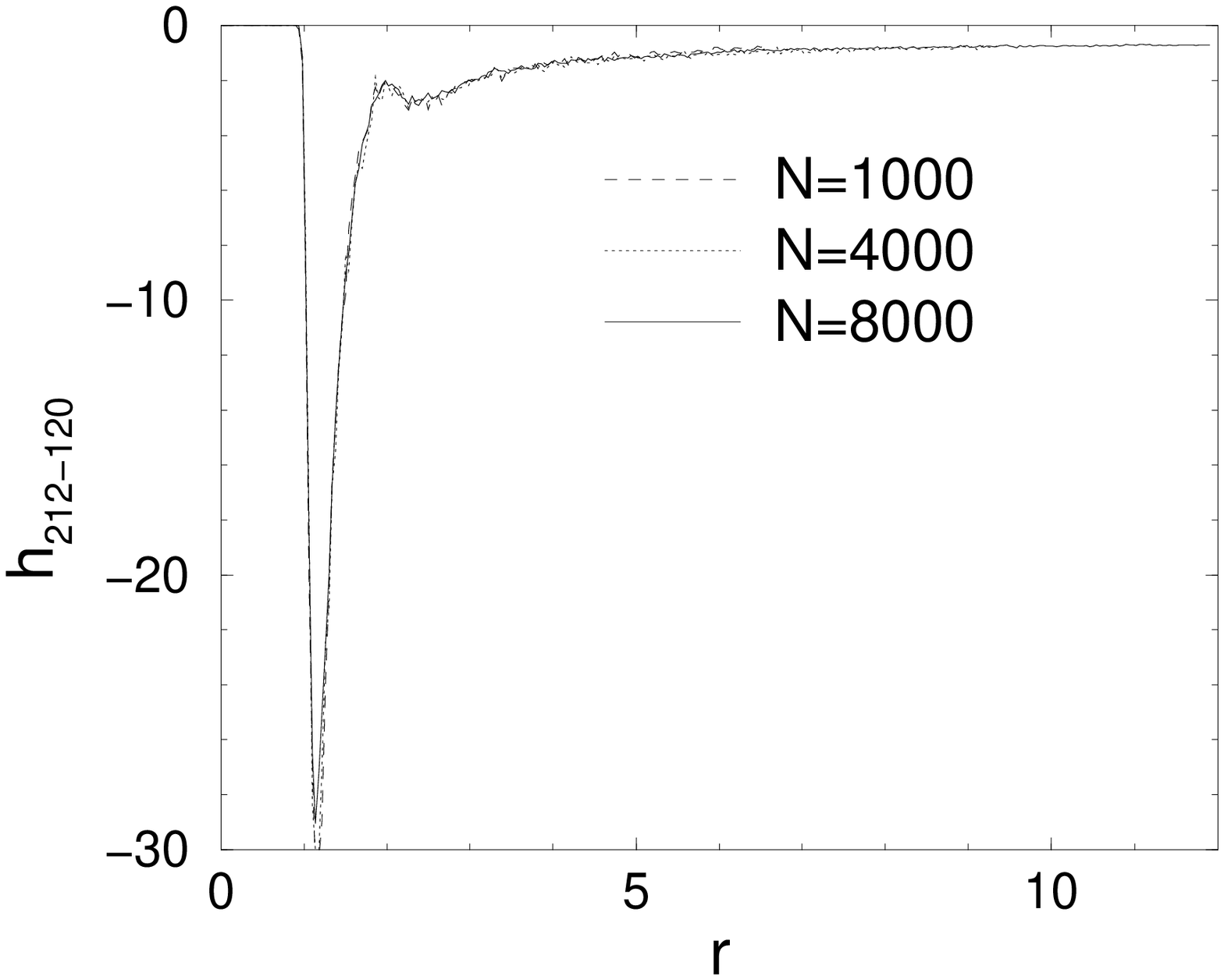}{55}{55} \fig{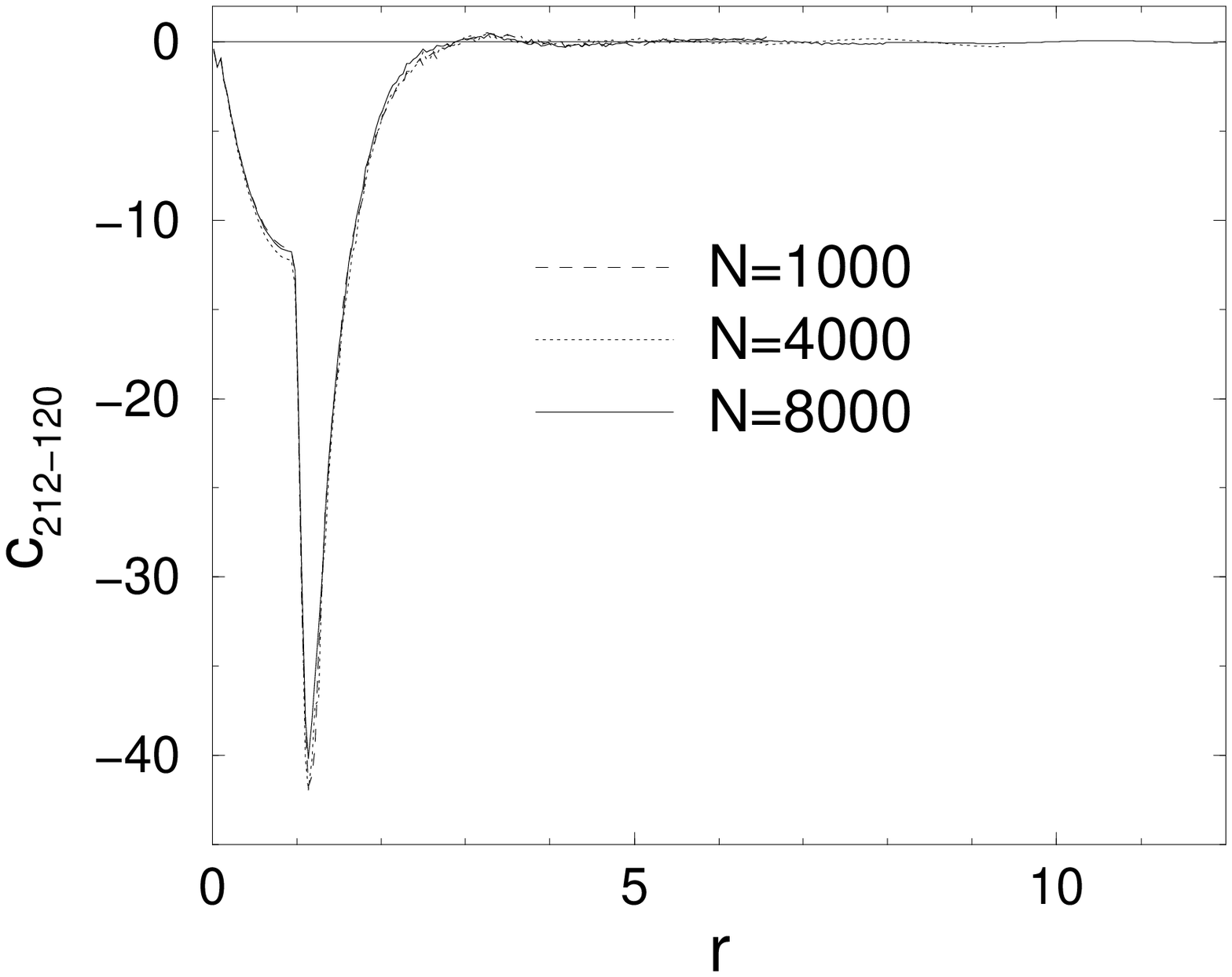}{55}{55}}

\vspace*{0.5cm}

{
\small
\noindent
{\bf Figure 7:}\\
Expansion coefficient with $l_1=2, l_2=2, m_1=1, m_2=-1, l=2$, and
$m=0$ of the total correlation function $h$ (left), and the 
direct correlation function (right) vs. molecular distance $r$ for
different system sizes $N$. 
Left Figure: Dashed line indicates an extrapolation with a $1/r$ power law
behavior. Inset shows the same data vs. $1/r$.
}

\section{Interfacial properties}

We turn to discuss briefly the interfacial properties in nematic liquid 
crystals. These are of great interest in the liquid crystal display 
technology~\cite{bahadur}. Therefore, a growing number of simulations 
are devoted to the study of model nematics at surfaces
\cite{stelzer3,stelzer4,stelzer5,palermo,miyazaki1,miyazaki2,demiguel4,demiguel5,doerr1,doerr2,emerson2,binger,xu,mike8,dennis}.
and interfaces~\cite{bates3,mike9,al-barwani,mcdonald,akino}  or in thin 
films~\cite{cleaver,chrzanowska,wall,mills,gruhn1,gruhn2,roij,dijkstra}.

The microscopic structure at surfaces can be quite 
complex~\cite{stelzer3,stelzer4,stelzer5,palermo}.
From a macroscopic point of view, the presence of surfaces or 
interfaces introduces essentially three new effective parameters:
the interfacial tension, the anchoring angle, and the anchoring energy.
The anchoring angle is the angle with respect to the surface which
the director preferably assumes close to the surface. The anchoring
energy is related to the force needed to twist the director out of
the anchoring angle.

The anchoring angle can be calculated from simulations in a straightforward 
way. Obtaining the anchoring energy is more difficult and time consuming. 
Allen and coworkers have devised different methods and applied them to 
systems of ellipsoids at hard, planar walls ~\cite{mike8,dennis}.
The interfacial tension can be computed from the anisotropy of the
pressure tensor close to the surface~\cite{mcdonald,mike10}, or in the
case of fluid-fluid interfaces from their undulations (capillary
waves)~\cite{akino}. Recently, a number studies have been devoted to the
interface between the nematic and the isotropic phase in
Gay-Berne fluids~\cite{bates3} and fluids of 
spherocylinders~\cite{al-barwani} or ellipsoids~\cite{mike9,mcdonald,akino}.
They reveal among other a rather intriguing capillary wave spectrum.
It reflects a complex interplay between the bare surface tension
and the elastic interactions in the nematic phase~\cite{akino}.

\section{Conclusions}

In materials science, systems of ellipsoids serve as model systems 
which exhibit generic properties of nematic liquid crystals. 
They are studied for the general insight they can give into the
physics of liquid crystals, and also because simulations of ellipsoids 
are computationally much cheaper than simulations of a comparable 
number of more realistic molecules. However, fluids of ellipsoids are also 
fascinating from a much more general point of view: They are simple systems 
where purely entropic effects generate an amazing variety of different 
structures, and lead to unusual long-range static and dynamic properties. 
Computer simulations are often the only way of investigating some of 
that behavior. 

\section*{Acknowledgements}

We have benefitted from discussions and interactions with 
N. Akino, M. P. Allen, K. Binder, G. Germano, and H. Lange. 
N. H. P. acknowledges financial support by the German Science
Foundation (DFG). The computer simulations of our group are
carried out mostly on the CRAY T3Es of the NIC in J\"ulich.

\end{document}